\documentclass[12pt,preprint]{aastex}
\shorttitle{A search for \lowercase{$o$}-C$_6$H$_4$ in CRL 618} \shortauthors{Widicus Weaver et al.}

\begin{document}

\title{A search for \lowercase{$ortho$}-benzyne (\lowercase{$o$}-C$_6$H$_4$) in CRL 618}

\author{Susanna L. Widicus Weaver}
\affil{Departments of Chemistry and Astronomy, University of Illinois at Urbana-Champaign, Urbana, IL  61801;
slww@uiuc.edu}

%\and
\author{Anthony J. Remijan}
\affil{National Radio Astronomy Observatory, Charlottesville, VA 22903; aremijan@nrao.edu}

\author{Robert J. McMahon}
\affil{Department of Chemistry, University of Wisconsin-Madison, Madison, WI 53706; mcmahon@chem.wisc.edu}

\and
\author{Benjamin J. McCall}
\affil{Departments of Chemistry and Astronomy, University of Illinois at Urbana-Champaign, Urbana, IL  61801;
bjmccall@uiuc.edu}

\begin{abstract}
Polycyclic aromatic hydrocarbons (PAHs) have been proposed as
potential carriers of the unidentified infrared bands (UIRs) and
the diffuse interstellar bands (DIBs). PAHs are not likely to form
by gas-phase or solid-state interstellar chemistry, but rather
might be produced in the outflows of carbon-rich evolved stars.
PAHs could form from acetylene addition to the phenyl radical
(C$_6$H$_5$), which is closely chemically related to benzene
(C$_6$H$_6$) and $ortho$-benzyne ($o$-C$_6$H$_4$). To date,
circumstellar chemical models have been limited to only a partial
treatment of benzene-related chemistry, and so the expected
abundances of these species are unclear.  A detection of benzene
has been reported in the envelope of the proto-planetary nebula
(PPN) CRL 618, but no other benzene-related species has been
detected in this or any other source. The spectrum of
$o$-C$_6$H$_4$ is significantly simpler and stronger than that of
C$_6$H$_5$, and so we conducted deep Ku-, K- and Q-band searches
for $o$-C$_6$H$_4$ with the Green Bank Telescope.  No transitions
were detected, but an upper limit on the column density of
8.4$\times10^{13}$ cm$^{-2}$ has been determined. This limit can
be used to constrain chemical models of PPNe, and this study
illustrates the need for complete revision of these models to
include the full set of benzene-related chemistry.

\end{abstract}


\section{Introduction}
Polycyclic aromatic hydrocarbons (PAHs) are large molecules with
carbon atoms arranged in five- or six-membered rings and are
thought to form from smaller carbon ring molecules such as benzene
(C$_6$H$_6$) and its derivatives.  PAHs are very stable against
photodissociation, and so could be present in interstellar and/or
circumstellar environments. The unidentified infrared bands (UIRs)
occur at frequencies characteristic of aromatic molecules, and so
PAHs have been suggested as potential UIR carriers. Likewise, PAHs
have been proposed as carriers of the diffuse interstellar bands
(DIBs). We refer the reader to a recent review of the DIB problem
by \cite{Sarre} and references therein for a discussion of PAHs
and their relation to the UIRs and DIBs. Significant laboratory
spectroscopic work has been dedicated to PAHs and other
benzene-related species to support astronomical observations, yet
none of these species has been unambiguously detected in space.
Only benzene has been tentatively detected toward the
proto-planetary nebula (PPN) CRL 618 \citep{Cernicharo2001a}.  One
vibrational band of benzene was observed, and a column density of
5$\times10^{15}$ cm$^{-2}$ and a kinetic temperature of 200 K were
determined.

CRL 618 is an evolved PPN with a central B0 star and an
ultra-compact H {\small II} region surrounded by a carbon-rich
molecular envelope \citep{Cernicharo2001b,Sanchez}. This source
has optical high velocity bipolar outflows \citep{Trammell} in
addition to a low velocity expanding torus of molecular emission
\citep{Sanchez}. CRL 618 has a rich molecular inventory including
a variety of hydrocarbons [see \cite{Pardo} and references
therein], but searches for biologically important molecules have
provided only upper limits \citep{Remijan}.

Very few observational constraints have been placed on the
chemical and physical properties of PPNe, especially regarding
benzene-related chemistry. Observational studies of the simplest
benzene derivatives are the essential first steps to understanding
the formation of PAHs in PPNe. The carbon-rich nature of CRL 618,
coupled with the benzene detection, makes it an excellent target
for initial benzene derivative searches. We have therefore
conducted a search for the benzene derivative $ortho$-benzyne
($o$-C$_6$H$_4$) in CRL 618 with the Green Bank Telescope (GBT)
Ku-, K-, and Q-band receivers. An overview of circumstellar
benzene-related chemistry, details of the observations, the
results of our search, and a discussion of these results are
presented below.

\section{Benzene-Related Circumstellar Chemistry}

PAHs have long been thought to be produced in carbon-star
outflows, as it seems unlikely that such large molecules could be
produced by gas-phase or grain-assisted chemistry within the
interstellar medium.  Although the PPN phase of stellar evolution
is typically very short, lasting only about ten thousand years,
the chemistry that takes place in the envelope of the
post-asymptotic giant branch (AGB) star leads to the formation of
complex hydrocarbons. PPN chemistry is steeply dependent on
density because circumstellar material undergoes energetic
processing, and formation mechanisms for large molecules are
likely a combination of radical-molecule and ion-molecule
reactions.  This is in contrast to low density interstellar
chemistry that is almost exclusively driven by ion-molecule
reactions. Both radical-molecule and ion-molecule reactions can
lead to the formation of benzene and related species, and such
benzene formation mechanisms have been included in PPN models
\citep{Redman,Woods}. However, no PPN model includes both the
radical- and ion-based benzene reaction networks.  This striking
deficiency in PPN models makes interpretation of observational
results quite difficult, and the radical- and ion-based chemical
mechanisms must be considered separately until more complete PPN
models are developed.

\begin{figure}[h!]
\includegraphics[angle=0, scale=0.60]{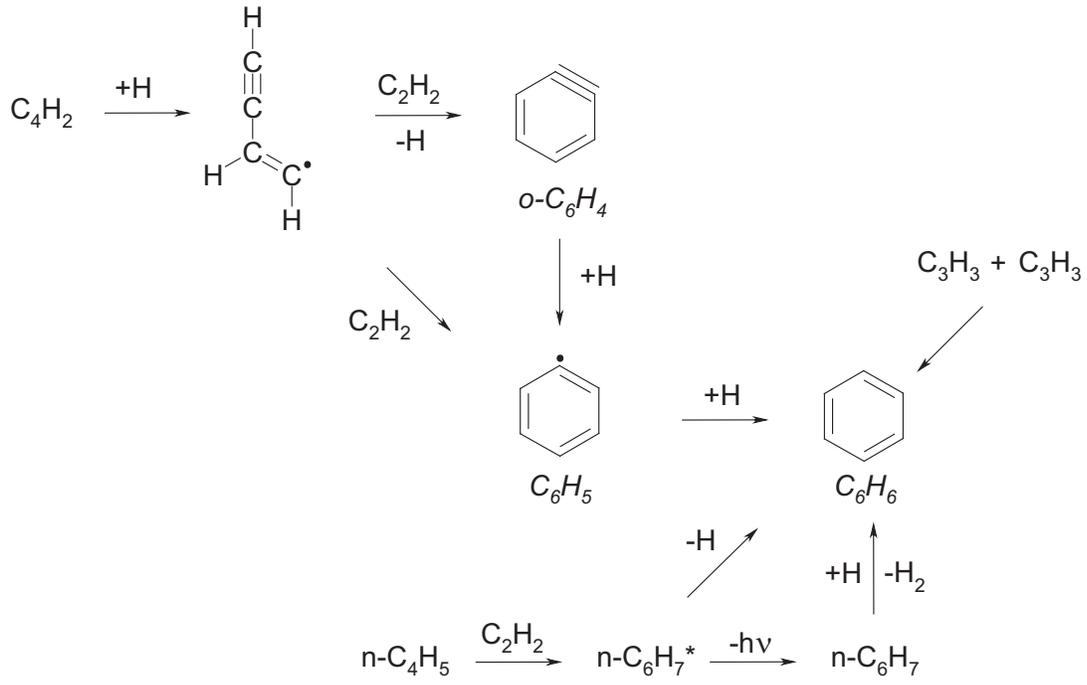}
\caption{Radical-based reaction scheme for PPNe benzene chemistry
based on \cite{Frenklach}.} \label{radical_rxns}
\end{figure}

The torus of a PPN is a high energy, high density environment, and
the chemistry is expected to be similar to that observed in
combustion \citep{Frenklach}. Radical-based mechanisms based on
combustion chemistry have been proposed for benzene and PAH
formation in the circumstellar shells of AGB stars
\citep{Cherchneff}, and a subset of these reactions was also used
to model the chemistry of clumps during PN evolution
\citep{Redman}. A summary of radical-driven benzene chemistry is
shown in Figure \ref{radical_rxns}.  Here, hydrocarbon radicals
react with acetylene (C$_2$H$_2$), and ring-closure forms
$o$-C$_6$H$_4$. Additional reactions lead to the phenyl radical
(C$_6$H$_5$) and benzene. Other routes to benzene include either
reaction of two C$_3$H$_3$ radicals or, alternatively, the
formation of straight-chain energetic C$_6$H$_7$ (n-C$_6$H$_7^*$),
which can then undergo ring-closure.

\begin{figure}[h!]
\includegraphics[angle=0, scale=0.60]{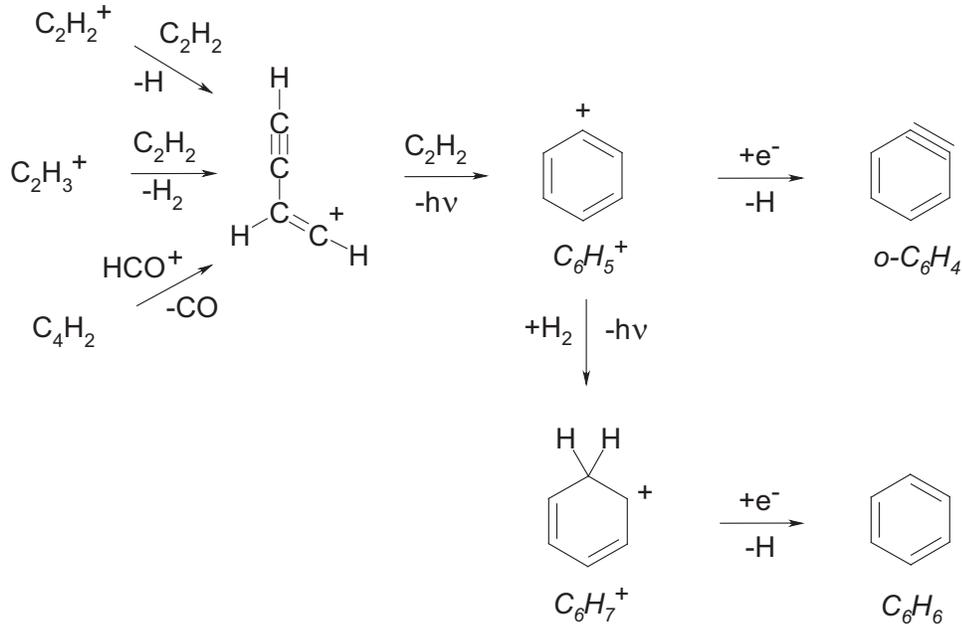}
\caption{Ion-based chemical scheme for PPNe benzene chemistry
based on  \cite{Woods}.} \label{ion_rxns}
\end{figure}

As a PPN evolves and its surrounding gas expands, the density
decreases and the material in the circumstellar envelope is
subjected to photoprocessing from the central star, leading to
ion-molecule chemistry.  An ion-molecule benzene formation
mechanism has been proposed for PPNe \citep{Woods}, and a summary
of this mechanism is shown in Figure \ref{ion_rxns}.  In this
network, hydrocarbon ions react with acetylene, and ring closure
forms C$_6$H$_5^+$. Additional hydrogenation and/or electron
recombination leads to $o$-C$_6$H$_4$, c-C$_6$H$_7^+$, and
ultimately C$_6$H$_6$. It should be noted that the original PPN
model investigated only the primary route to benzene shown in
Figure \ref{ion_rxns} and did not include the $o$-C$_6$H$_4$
formation reaction \citep{Woods}.

\begin{figure}[h!]
\includegraphics[angle=0, scale=0.60]{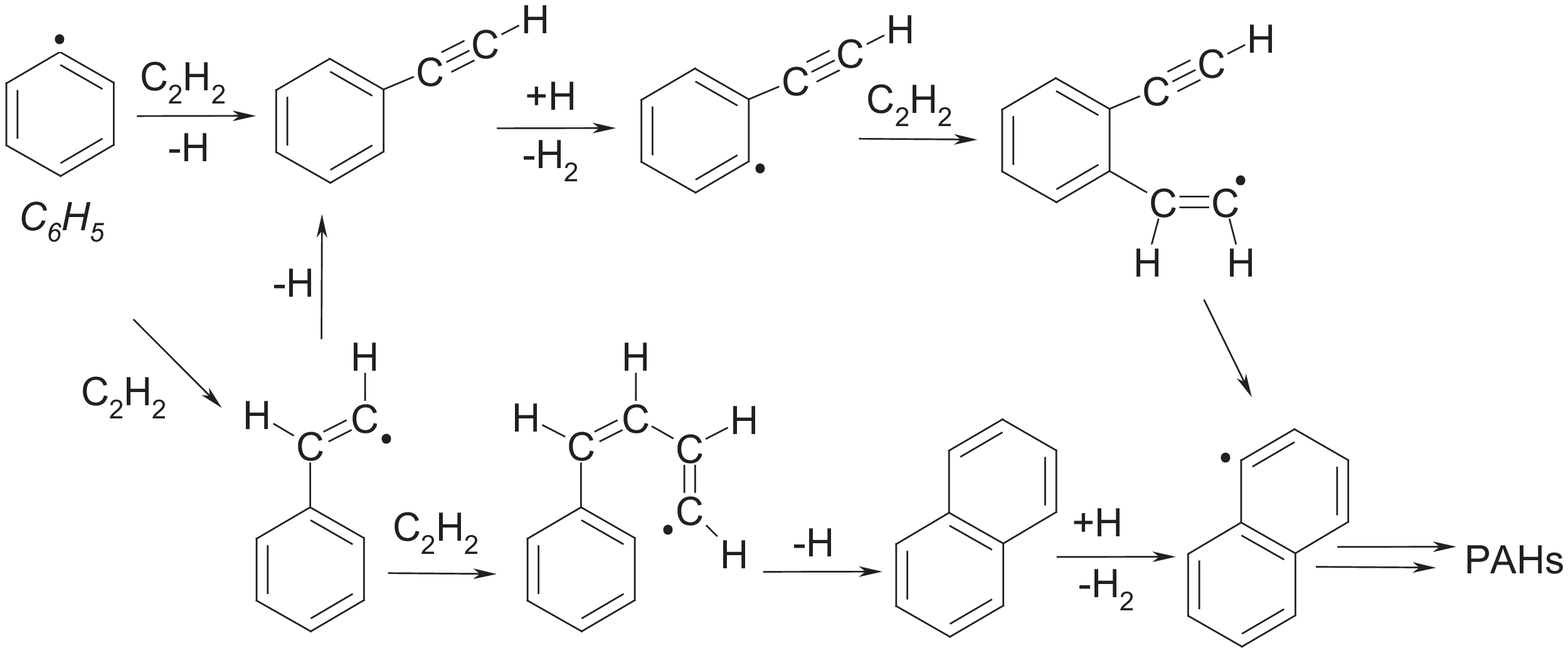}
\caption{Potential PAH formation routes in circumstellar shells
\citep{Frenklach,Cherchneff}.} \label{PAH_rxns}
\end{figure}

One of the benzene derivatives formed in these reactions,
C$_6$H$_5$, is a suggested precursor to PAHs in circumstellar
environments \citep{Frenklach,Cherchneff}. As is illustrated in
Figure \ref{PAH_rxns}, C$_6$H$_5$ can undergo a series of
radical-molecule reactions involving acetylene to produce a
naphthalene-like species (C$_{10}$H$_8$)  via ring closure.
Subsequent reactions of this nature can lead to larger PAHs.

Studies of other high energy environments indicate that C$_6$H$_5$ and $o$-C$_6$H$_4$ may well coexist with benzene in
CRL 618 if circumstellar chemistry is similar to combustion or plasma chemistry. The reaction network of
\cite{Frenklach} is based on soot production mechanisms in hydrocarbon flames, and theoretical studies show that
benzene unimolecular decomposition leads to C$_6$H$_5$ and $o$-C$_6$H$_4$ during combustion \citep{Mebel}. Benzene
electrical discharges also produce C$_6$H$_5$ and $o$-C$_6$H$_4$ \citep{McMahon}. There are no reliable predictions of
C$_6$H$_6$, C$_6$H$_5$, and $o$-C$_6$H$_4$ abundances in PPNe because of the partial treatment of their chemistry in
models, but observations of C$_6$H$_5$ and $o$-C$_6$H$_4$ would provide important limits for future modelling studies.
We therefore began a search for these species in CRL 618.

C$_6$H$_6$ has no permanent dipole moment and can therefore only be studied in the infrared, but both C$_6$H$_5$ and
$o$-C$_6$H$_4$ can be probed by radioastronomical techniques. The rotational spectra of $o$-C$_6$H$_4$ and C$_6$H$_5$
have  been obtained in the laboratory \citep{Brown,Robertson,Kukolich,McMahon}. Both species are carbon ring, planar,
near-oblate asymmetric rotors with C$_{2v}$ symmetry along their $b$ inertial axes. The calculated dipole moment of
$o$-C$_6$H$_4$ is 1.38 D \citep{Kraka} and that of C$_6$H$_5$ is 0.9 D \citep{McMahon}. The unpaired electron in
C$_6$H$_5$ leads to hyperfine splitting of the lower rotational states \citep{McMahon}, and so there are many
rotational transitions to sample observationally. However, the larger dipole moment coupled with the lack of hyperfine
splitting yields much stronger lines for $o$-C$_6$H$_4$ and makes this species a more likely candidate for detection if
the abundances are similar.

\section{Observations}

Observations of $o$-C$_6$H$_4$ were conducted with the
NRAO\footnote{The National Radio Astronomy Observatory is a
facility of the National Science Foundation, operated under
cooperative agreement by  Associated Universities, Inc.} 100 m
Robert C. Byrd GBT between 2006 September 4 - 2007 January 28
using the Ku-band (12 - 15.4 GHz), K-band (18 - 22.5 and 22 - 26.5
GHz,) and Q-band (40 - 48 GHz) receivers. The eight
intermediate-frequency (IF), 200 MHz, three-level GBT spectrometer
configuration mode was used, providing four 200 MHz frequency
bands with a channel separation of 24.4 kHz in two polarizations.
The assumed CRL 618 J2000 pointing position and LSR source
velocity were $\alpha$= 04$^h$42$^m$53$^s$.7, $\delta$ =
+36$^o$06$'$53$''$.0 and -27.5 km s$^{-1}$, respectively. Data
were acquired in the OFF-ON position-switching mode.  A scan
included two-minute integrations for each position beginning with
the OFF-source position, which was located 60$'$ East  in azimuth
of the ON-source position. Antenna temperatures with estimated
20\% uncertainties were recorded on the $T_A^*$ scale
\citep{Ulich}.  The GBT half-power beamwidths are given by
$\theta_b$ = 740\arcsec/$\nu$(GHz). Dynamic pointing and focusing
corrections were applied and observations of the quasar 0359+509
were used to adjust the zero points every two hours or less.
Observations from multiple nights and both polarizations were
averaged for each frequency window, and the data were Hanning
smoothed over three channels with the GBDish data reduction
program.  The resultant Q-band spectrum is shown in Figure
\ref{figure:spectrum}.

Table 1 lists the $o$-C$_6$H$_4$ rotational transitions in the
observed frequency windows. The transition quantum numbers,
transition rest frequency ($\nu$), the Einstein $A$ coefficient of
the transition times the upper state degeneracy ($Ag_u$), the
energy of the upper level ($E_u$), the observed 1-$\sigma$ RMS
level ($T_{MB}$), and the beam efficiency ($\eta$) are listed in
the first six columns. The transition frequencies and intensities
are from the Cologne Database for Molecular Spectroscopy
\citep{Muller}.  The d.rms routine in GBDish was used to calculate
the RMS level in the $T_A^*$ scale for each spectral window.  The
RMS level is equal to the standard deviation of the noise in a
line-free region and was calculated after the Hanning smoothing
was applied. These values were then converted to the $T_{MB}$
scale by the relationship $T_{MB}$ = $T_A^*$/$\eta$. The value of
$\eta$ was derived from a fit to the \cite{Ruze} formulation, as
is outlined in Equation (2) of \cite{Hollis}.

\section{Results and Discussion}
\begin{figure}[h]
\includegraphics[angle=0, scale=0.6]{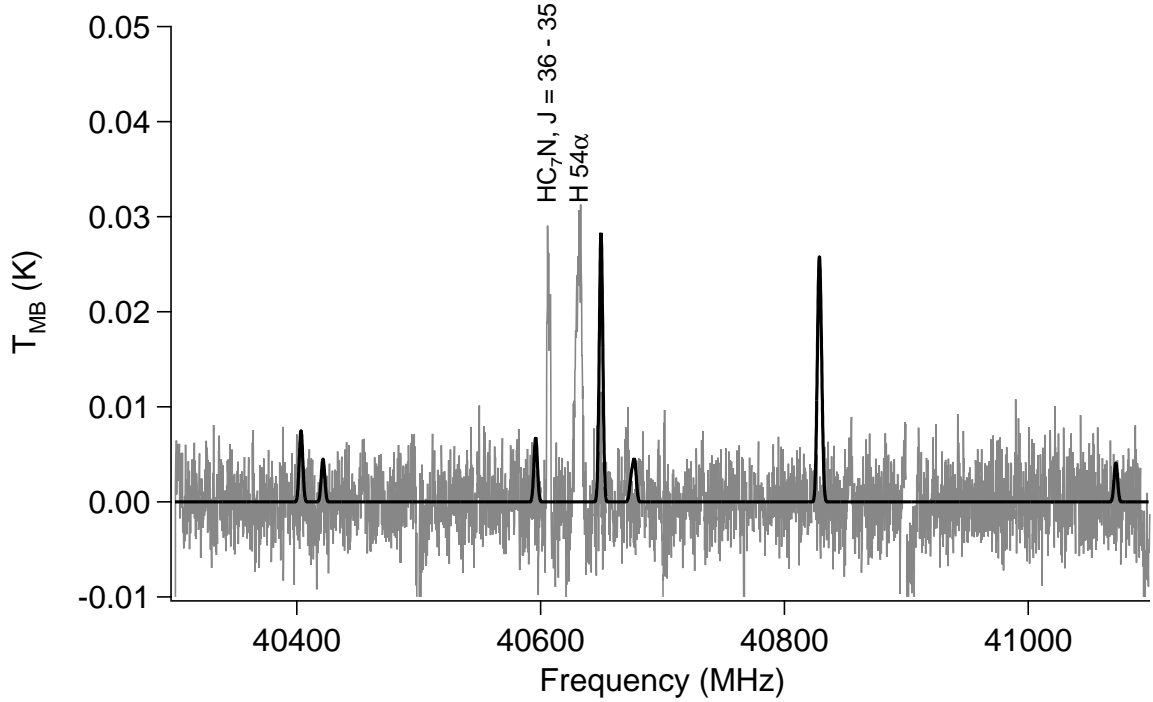}
\caption{Observed CRL 618 Q-band spectrum (grey) overlaid with a
simulation of the $o$-C$_6$H$_4$ spectrum (black) at a column
density of 10$^{16}$ cm$^{-2}$, a rotational temperature of 200 K,
and a source size of 10\arcsec.  The observed spectrum is
comprised of four 200 MHz windows and has been smoothed to a
resolution of 200 kHz.  The features at 40500 and 40900 MHz arise
from noise at the edges of the spectral windows.}
\label{figure:spectrum}
\end{figure}

No spectral features associated with $o$-C$_6$H$_4$ were detected
during this search, and there are no unidentified spectral
features in any passband. Figure \ref{figure:spectrum} shows a
plot of the observed CRL 618 Q-band spectrum overlaid by a
predicted $o$-C$_6$H$_4$ spectrum at a column density of 10$^{16}$
cm$^{-2}$, a rotational temperature of 200 K, and a source size of
10\arcsec. If these parameters are representative of
$o$-C$_6$H$_4$, the blended 6$_{0,6 }-5 _{1 ,5 }$ and 6$_{1,6 }-5
_{0 ,5 }$ transitions at 40828.1686 and  40829.9929 MHz,
respectively, should have been easily detected at the 30 mK level.

The RMS levels reached during the observations allow calculation
of the  $o$-C$_6$H$_4$ column density upper limit in CRL 618, and
the 3-$\sigma$ upper limits determined from each observed
transition are presented in Table \ref{Table 1}. These column
density upper limits were calculated using the following
expression, adapted from equation (1) of Nummelin et al. (1998):

\begin{equation}\label{Integrated_Intensity}
  {{N_T}={\int_{-\infty}^{\infty}{T_{b}}{dv}{\:\:}{\frac{8{\pi}{k}{\nu^2}}{hc^3}}{\frac{Q(T_{rot})}{A
{g_u}}}\:{e^{{E_u}/{kT_{rot}}}}}}
\end{equation}
\smallskip

\noindent where $N_T$ is the beam averaged total column density,
$\int_{-\infty}^{\infty}{T_{b}}{dv}$ is the transition integrated
intensity, $Q(T_{rot})$ is the rotational partition function, and
$T_{rot}$ is the molecular rotational temperature. Since no lines
were observed, the integrated intensity was approximated as that
of a line with a peak intensity at the 3-$\sigma$ RMS level,
$T_{RMS}$, and an assumed linewidth ${\Delta}{v}$.  The value of
$T_{RMS}$ was determined by $T_{RMS}$=3$T_{MB}$/$\surd{n}$, where
$n$ is the number of channels across the linewidth ${\Delta}{v}$.
The 1/$\surd{n}$ factor does not include a complete statistical
treatment of the assumed Gaussian lineshape, but it does
approximate the overestimation of the RMS level from spectral
oversampling. Examination of previously reported upper limit
calculations indicates that, in nearly all cases, this factor is
either neglected entirely, or at the very least not discussed
explicitly. We find this to be a gross omission for instruments
such as the GBT spectrometer, where the channel width is often
significantly smaller than the linewidth. From the $T_{RMS}$
values, correction for beam dilution gave $T_b$ through the
expression $T_{b}$=$T_{RMS}/B$, where the beam filling factor,
$B$, was calculated from the source size, $\theta_s$, and the beam
size, $\theta_b$, by the relationship
$B=\theta_s^2/[\theta_s^2+\theta_b^2]$.   \noindent The integrated
intensity was then approximated as
$\int_{-\infty}^{\infty}T_b\:dv=1.064\:T_{b}\:\Delta{v}$, where
the 1.064 factor arises from the assumed Gaussian lineshape.  We
have included this factor for completeness, although it is likely
minimal compared to the uncertainty in $T_b$.

The calculated 3-$\sigma$ $o$-C$_6$H$_4$ upper limits for CRL 618
are presented in Table \ref{Table 1}. These calculations required
assumptions for the values of $T_{rot}$, ${\Delta}{v}$, and
$\theta_s$.  The linewidth was assumed to be that observed for
other identified lines, 10 km s$^{-1}$. Determination of an
appropriate temperature and source size was less straightforward,
as there is much discrepancy in the literature regarding these
values. The kinetic temperature derived for benzene is 200 K
\citep{Cernicharo2001a}, but an IRAM 30-meter millimeter line
survey indicates temperatures of 250 - 275 K for the torus and 30
K for the circumstellar shell \citep{Pardo}. It is expected that
$o$-C$_6$H$_4$ would have very similar physical properties to
those of benzene in this source, and so a temperature of 200 K was
assumed for the upper limit calculations.

There is similarly contradicting information regarding the CRL 618
source size. \cite{Pardo} found torus and circumstellar shell
sizes of 1.5\arcsec{} and 3.0 - 4.5\arcsec, respectively, while
millimeter interferometric observations with a $\sim$5\arcsec{}
beam gave source sizes $\leq$10\arcsec{} for molecules expected to
trace the extended envelope \citep{Remijan}. A molecule such as
$o$-C$_6$H$_4$ is likely to be present in the torus where higher
densities shield it from photodissociation, but such a species
could also be present in more extended regions if readily formed
by ion-molecule chemistry. The GBT beam is $\geq$16\arcsec{}, and
so our observations probed both the torus and the extended
envelope of CRL 618.  A source size of 10\arcsec{} was assumed for
the upper limit calculations, as this includes the entire
molecular envelope and therefore all possible regions of emission.

As is shown in Table \ref{Table 1}, the most strict $o$-C$_6$H$_4$
column density upper limit determined from these observations is
8.4$\times10^{13}$ cm$^{-2}$. The observed benzene column density
is $5\times10^{15}$ cm$^{-2}$ \citep{Cernicharo2001a}, and
\cite{McMahon} report a C$_6$H$_5$ column density upper limit in
CRL 618 of 4 $\times 10^{15}$ cm$^{-2}$.  The $o$-C$_6$H$_4$ upper
limit is therefore much lower than the limits for the related
species. Given the incomplete nature of the chemical models,
however, the significance of this limit is unclear.

Further interpretation of this limit will require additional
observational and modelling studies, as this work highlights the
nearly total lack of information regarding the physical and
chemical properties relevant to benzene-related chemistry in
circumstellar environments. Interferometric observations would
prove quite useful, as such studies would probe the spatial
distribution of molecules in the torus and extended envelope,
eliminating the uncertainty in beam dilution effects. Laboratory
measurements of the millimeter and submillimeter spectra of
$o$-C$_6$H$_4$ would aid these observations.  We also strongly
encourage complete revision of existing PPN models to investigate
both the radical- and ion-based benzene chemical networks for
comparison to $o$-C$_6$H$_4$, C$_6$H$_5$, and C$_6$H$_6$
observations.  Only a combination of further observations,
modelling, and laboratory measurements of benzene derivatives will
lead to full understanding of PAH formation mechanisms in PPNe.

\acknowledgments We would like to thank the NRAO and the GBT
support staff, especially Carl Bignell.  Support for SLWW and BJM
was provided by the NSF CAREER award (NSF CHE-0449592) and the
UIUC Critical Research Initiative program. Support for RJM was
provided by NSF-0412707. We are grateful to Matthew Redman and
Paul Woods for providing additional details about their chemical
models.  We would also like to thank Michael Remijan for
programming support and development.
%\clearpage

\clearpage

\begin{table}[h!]
\center \caption{Summary of $o$-C$_6$H$_4$ Observations Toward CRL
618\label{Table 1}}
\begin{small}
\begin{tabular}{cccrccc}
\tableline\tableline
\multicolumn{7}{c}{\hspace{0.1pt}}\\
                                       &          &                &       &           &        & $N_T$     \\
J$'_{K_a',K_c'}-$J$''_{K_a'',K_c''}$   & $\nu$    & $Ag_u$         & $E_u$ & $T_{MB}$  & $\eta$ & Upper Limit     \\
                                       & (MHz)    & (s$^{-1}$)     & (K)   &  (mK)     &        &   (cm$^{-2}$)     \\
\multicolumn{7}{c}{\hspace{0.1pt}}\\
\tableline
\multicolumn{7}{c}{\hspace{0.1pt}}\\
4 $_{2,2 }-4 _{1 ,3 }$ & 13250.8230  &   6.08$\times10^{-7}$   &  5.28& 1.48 & 0.92 & 3.47$\times10^{15}$\\
3 $_{3,1 }-3 _{2 ,2 }$ & 13680.0050  &   3.59$\times10^{-7}$   &  3.70& 1.76 & 0.92 & 6.87$\times10^{15}$\\
6 $_{5,1 }-6 _{4 ,2 }$ & 14608.9021  &   1.28$\times10^{-6}$   & 12.37& 1.45 & 0.91 & 1.59$\times10^{15}$\\
3 $_{1,2 }-3 _{0 ,3 }$ & 14619.6969  &   2.40$\times10^{-7}$   &  2.96& 1.45 & 0.91 & 8.22$\times10^{15}$\\
2 $_{0,2 }-1 _{1 ,1 }$ & 14753.9974  &   4.23$\times10^{-7}$   &  1.19& 1.45 & 0.91 & 4.62$\times10^{15}$\\
3 $_{0,3 }-2 _{1 ,2 }$ & 21770.7375  &   1.60$\times10^{-6}$   &  2.26& 2.47 & 0.86 & 1.79$\times10^{15}$\\
4 $_{2,3 }-4 _{1 ,4 }$ & 21963.3591  &   7.93$\times10^{-7}$   &  4.67& 2.47 & 0.86 & 3.63$\times10^{15}$\\
11$_{8,3 }-11_{7 ,4 }$ & 22142.0552  &   5.35$\times10^{-6}$   & 37.44& 2.18 & 0.85 & 5.28$\times10^{14}$\\
9 $_{5,4 }-9 _{4 ,5 }$ & 22143.1890  &   4.10$\times10^{-6}$   & 24.25& 2.18 & 0.85 & 6.60$\times10^{14}$\\
3 $_{1,3 }-2 _{0 ,2 }$ & 22216.3288  &   2.88$\times10^{-6}$   &  2.26& 2.18 & 0.85 & 8.72$\times10^{14}$\\
7 $_{5,3 }-7 _{4 ,4 }$ & 23578.9636  &   5.35$\times10^{-6}$   & 15.55& 2.51 & 0.84 & 5.56$\times10^{14}$\\
2 $_{2,1 }-1 _{1 ,0 }$ & 24109.5097  &   1.40$\times10^{-6}$   &  1.77& 1.76 & 0.84 & 1.42$\times10^{15}$\\
3 $_{1,2 }-2 _{2 ,1 }$ & 24842.4030  &   8.22$\times10^{-7}$   &  2.96& 2.13 & 0.83 & 2.90$\times10^{15}$\\
8 $_{2,6 }-8 _{1 ,7 }$ & 40403.3815  &   1.64$\times10^{-5}$   & 16.25& 6.42 & 0.64 & 4.19$\times10^{14}$\\
8 $_{3,6 }-8 _{2 ,7 }$ & 40421.4144  &   9.83$\times10^{-6}$   & 16.25& 6.42 & 0.64 & 6.98$\times10^{14}$\\
5 $_{1,4 }-4 _{2 ,3 }$ & 40595.9260  &   1.40$\times10^{-5}$   &  6.62& 5.92 & 0.63 & 4.37$\times10^{14}$\\
22$_{13,9}-22_{12,10}$ & 40649.5538  &   1.13$\times10^{-4}$   &137.89& 5.92 & 0.63 & 8.40$\times10^{13}$\\
7 $_{1,6 }-7 _{0 ,7 }$ & 40674.4124  &   5.21$\times10^{-6}$   & 11.45& 5.92 & 0.63 & 1.19$\times10^{15}$\\
7 $_{2,6 }-7 _{1 ,7 }$ & 40677.1133  &   8.69$\times10^{-6}$   & 11.45& 5.92 & 0.63 & 7.16$\times10^{14}$\\
6 $_{0,6 }-5 _{1 ,5 }$ & 40828.1686  &   4.05$\times10^{-5}$   &  7.23& 6.21 & 0.63 & 1.59$\times10^{14}$\\
6 $_{1,6 }-5 _{0 ,5 }$ & 40829.9929  &   2.43$\times10^{-5}$   &  7.23& 6.21 & 0.63 & 2.64$\times10^{14}$\\
3 $_{3,0 }-2 _{2 ,1 }$ & 41071.9441  &   8.57$\times10^{-6}$   &  3.74& 9.40 & 0.63 & 1.12$\times10^{15}$\\

\tableline
\end{tabular}
\end{small}
\end{table}

\clearpage

\end{document}